\documentclass{aa}

\usepackage{graphics}

\newcommand{\kms}{km~s$^{-1}$}
\newcommand{\ha}{H$\alpha$}
\newcommand{\hb}{H$\beta$}
\newcommand{\simgt}{\lower.5ex\hbox{$\;\buildrel>\over\sim\;$}}
\newcommand{\simlt}{\lower.5ex\hbox{$\;\buildrel<\over\sim\;$}}
\begin{document}

\thesaurus{11.01.2; 11.14.1; 11.19.1}
\title{A search for variability in Narrow Line Seyfert 1 Galaxies.
II. New data from the Loiano monitoring programme}

\author{M. Ester Giannuzzo \inst{1,}\inst{2} 
%
%
, Marco Mignoli \inst{3}, Giovanna M. 
Stirpe \inst{3} \and Andrea Comastri \inst{3}}

\offprints{M. E. Giannuzzo}

\institute{
Department of Physical Sciences, University of Hertfordshire, 
College Lane,\\ 
Hatfield HERTS AL10 9AB, UK -- email: ester@star.herts.ac.uk\\
\and
Dipartimento di Astronomia, Universit\`a degli Studi di Bologna, Via 
Zamboni 33, I--40126 Bologna, Italy\\
\and
Osservatorio Astronomico di Bologna, Via Zamboni 33, I--40126 Bologna, Italy}

\date{Received 1997 / Accepted}

\titlerunning{Variability in Narrow Line Seyfert 1 Galaxies. II}
\authorrunning{M. E. Giannuzzo et al.}
 
  \maketitle

   \begin{abstract}

We present part of the results from a spectroscopic
monitoring programme on a sample of 
AGNs, relative to Narrow
Line Seyfert 1 Galaxies: following the idea that
Balmer-line variability can help discriminate among the possible models
for these objects, and on
the basis of a first monitoring project with long time-scales, we now add 
to the previous database new variability results, both on short and
long time-scales.
The comparison with a similar data-set of normal Seyfert 1's suggests now
a statistically different behaviour of NLS1s with regard to the 1-year
variability properties, and a clearer trend of NLS1s towards weaker
variability is suggested by the comparison between the 1-month variations
of NLS1s and those of the typical Seyfert 1 NGC 5548. Although these results
do not take into account the possible role of the relative contribution of 
narrow components to the line fluxes, and of the 
continuum variability amplitudes (both of which can in principle
be different in the two classes of objects), they are consistent with
the model of a BLR which is on average larger in NLS1s.

\keywords{Galaxies: active -- Galaxies: nuclei -- Galaxies: Seyfert}
   
\end{abstract}


\section{Introduction}

The widespread property of variability of the continuum and
broad lines emitted by Active Galactic Nuclei has been used 
in recent years to obtain constraints on the structure and kinematics
of the Broad Line Region, through intensive spectral monitoring and 
the use of 
reverberation mapping techniques (Blandford \& McKee \cite{blandford}, 
Peterson \cite{peterson},
Robinson \cite{r94}, Alloin et al. \cite{alloin}): the inferred sizes 
of the BLR
range from a few light days to about a light
month, and therefore direct information on the region through imaging
techniques is not available.

Narrow Line Seyfert 1 Galaxies (Osterbrock \& Pogge \cite{osterbrock}) are
a sub-class of Seyfert 1's located at the lower end of the broad line 
width distribution: the FWHM of the permitted lines does
not exceed 1000-1500~\kms\ in these objects. In most other
properties NLS1s are similar to `normal' Seyfert 1's: they have 
different widths of 
permitted and forbidden lines, intense \ion{Fe}{ii} lines and 
high ionization transitions, normal
line ratios and average luminosities.
They present, on average, lower equivalent widths compared to Seyfert 1's
(Osterbrock \& Pogge \cite{osterbrock}, Goodrich
\cite{goodrich}), but again this is the extension to low FWHM of a trend 
observed throughout the whole Seyfert 1 population.
NLS1s are efficiently found in soft X-ray
selected Seyfert 1 samples, in which they represent $\sim$ 16-50\% of all  
objects (Stephens \cite{stephens}, Puchnarewicz et al. \cite{puchna}), 
compared to the 
$\sim$ 10\% found in optically-selected samples. Boller et al. (\cite{boller})
found that NLS1s  
have generally much steeper soft X-ray (ROSAT) continuum slopes than 
normal Seyfert 1's, and present in a few cases rapid soft X-ray variability.
The 2-10 keV spectra have also been found to be steeper in NLS1s than in
other Seyfert 1's (Brandt et al. \cite{brandt}).

The line-of-sight component of the velocity of the
broad line emitting gas may be unusually small for several reasons, among
which a lower mass of the central black hole, projection effects
due to gas motions in a plane (e.g. that of a disk)
observed almost pole-on, a larger BLR size,
a lack or obscuration of the inner, high velocity regions of the BLR.
The explanation of this behaviour
can therefore provide insight on the more general problem posed by the great
diversity among broad emission line profiles. 

Permitted line variability, being directly related to the size, geometry
and kinematics of the Broad Line Region, is potentially
very useful to discriminate among some of the mentioned hypotheses: a 
lack of widespread variability in NLS1s would in fact suggest  
that the visible BLR is located at a higher distance from the central
engine than in
normal Seyfert 1's; similar variability properties in NLS1s and Seyfert 1's
could instead imply a smaller central mass or an anisotropic 
kinematic structure for the BLR. 

Since no NLS1 has ever been intensively monitored, and virtually no
information was available on the presence of variability in this
class of objects when the project started, we performed
a preliminary simple observational programme aimed at 
determining whether variability is a common characteristic of NLS1s as a 
class (Giannuzzo \& Stirpe \cite{gs95}, Giannuzzo \& Stirpe 
\cite{gs96} -- hereafter 
Paper I): we adopted a statistical approach, by evaluating the optical line
flux percentage variations on a time-scale of 1 year for a sample of NLS1s,
and comparing our results with those of an existing similar data-set
of normal Seyfert 1's. We did not find evidence for a weaker or less common
variability in Narrow Line Seyfert 1's, and discussing the competing models
we concluded that a larger BLR or a smaller black hole mass could be the most
probable explanations; we stressed that
a short-term monitoring program could help 
discriminate between the two possibilities. This kind of observations could
also address the problem of the generality of reverberation
mapping results, since until now only AGNs with high 
variability and normal line
profiles have been monitored, perhaps discriminating against different
types of BLRs (Robinson \cite{r95}).

In this work the observations reported in Paper I are added to new data
on NLS1s from a spectroscopic monitoring programme, conducted at the
Bologna Observatory on a larger sample 
of AGNs, which lasted a few years:
this following step of the project aims therefore at collecting
information on the variability properties of a greater number of
objects on different time-scales, though not in much detail.
We have thus enlarged the previous data-set on 1-year percentage
variations of the optical lines (this time both \hb\ and \ha\ have 
been considered),
allowing in this way a more accurate statistical
comparison than before; furthermore, we evaluated the short-term 
variations of the line fluxes with the aim to compare them, in the case of \ha,
with those of some monitored Seyfert 1's, while in
the case of \hb\ we used as a comparison object the
best studied Seyfert 1, NGC 5548, the light
curve of which has been well sampled for several years (Korista et al.
\cite{korista} and references therein).

Section 2 describes the observations and data reduction and analysis,
section 3 presents our results and the statistical comparisons between
NLS1 and Seyfert 1 variability properties; finally, section 4 includes
a discussion on the results and our conclusions.

   \begin{figure*}

\resizebox{\hsize}{!}{\includegraphics{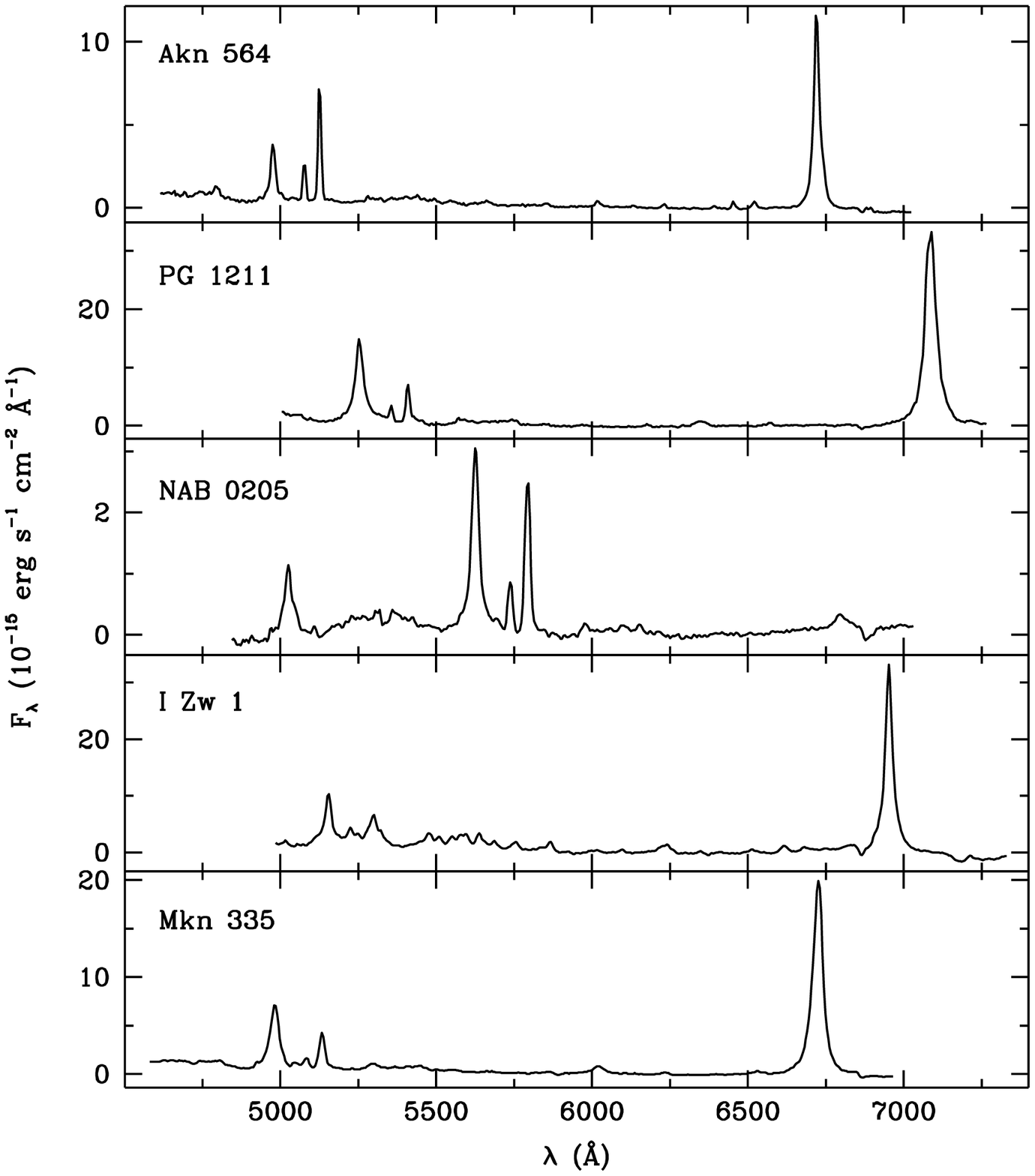}}
\caption{Representative continuum-subtracted
spectra of the monitored Narrow Line Seyfert 1 objects.}
\label{spettri}

    \end{figure*}

\section{Observations, data reduction and analysis}

The sample of AGNs chosen for the monitoring campaign includes 5 objects
classified (sometimes `marginally') as Narrow Line Seyfert 1's, 2 of 
which were included also in the sample observed at ESO (Paper I):
more specifically, the NLS1 group includes the prototype of the class, 
I Zwicky 1 (e.g. Halpern \& Oke \cite{halpern}), Akn 564 (Osterbrock \&
Shuder \cite{shuder}), NAB 0205+024, Mkn 335 and PG 1211+143 (classified 
by us on the basis of
published spectra according to the Osterbrock \& Pogge \cite{osterbrock}
criteria). The last two objects are considered as `marginal' because
their permitted line FWHM is slightly larger than the $\sim$ 1000~\kms\
limit, being 1200~\kms\ in Mkn 335 (unpublished data collected by the
authors) and 1600~\kms\ in PG 1211+143 (Stirpe \cite{stirpe}), but in
both cases the narrow line had been subtracted, which increased the
measured line FWHM; furthermore, they share other properties with
typical NLS1s, such as intense \ion{Fe}{ii} emission and weak forbidden
lines.
A sample continuum-subtracted
spectrum for each of the observed NLS1s is plotted in
Figure~\ref{spettri}.

The observations were carried out at the Cassini 1.52 m telescope operated by
the Bologna Observatory, using two different spectrographs. A log of the
observations is given in Table~\ref{log}. 
In the first part of the monitoring campaign
(October 1991 -- March 1994) the spectra were obtained with a Thomson~7882
CCD (576$\times$384 pixels) mounted on a Boller \& Chivens spectrograph.
The 350 l/mm grating used at first order gives a coverage of 2500 \AA\ with a
resolution of 4.5 \AA/pixel. The grating angle was adjusted for each object in
order to cover, when possible, both the \ha\ and \hb\ spectral regions.
In the summer of 1994, a new all-transmitting spectrograph named \hbox{BFOSC}
(Bologna Faint Object and 
Spectrograph Camera) replaced the Boller \& Chivens.
The new instrument, working at the Cassegrain focus of the telescope
as a focal reducer, is based on a EFOSC concept (Enard \& Delabre 
\cite{enard}).
The observations after July 1994 were therefore carried out 
with a Thomson
coated CCD (1024$\times$1024) mounted on BFOSC, with a projected pixel size of
0.56 arcsec. The used grism gives a larger but fixed spectral range
(4000--7850 \AA) and a resolution of 4 \AA/pixel. 

\begin{table*}

\caption[]{Log of Loiano and ESO (marked with `(E)')
observations used throughout this work.}
\label{log}

\begin{flushleft}
\begin{tabular}{lccccccccc}
\hline
\hline
\\
Galaxy & $\alpha$ (1950) & $\delta$ (1950) & $m_V$ & $z$ & Date of 
observation & Julian date& Integration time \\
& & & & & & (JD -- 2400000) & (minutes) \\
\\
\hline
\\
Mkn 335 & 
	    00 03 45.2 & +19 55 29 & 13.85 & 0.025 & 1991 Oct 16 & 48546.4 & 30 \\
 & 
	     & & & & 1992 Aug 1 & 48836.5 & 30 \\
 & 
	     & & & & 1992 Sep 1 & 48867.4 & 30 \\
 & 
	     & & & & 1993 Aug 11 & 49211.5 & 90 \\
 & 
	     & & & & 1993 Aug 20 & 49220.5 & 60 \\
 & 
	     & & & & 1994 Aug 8 & 49573.5 & 40 \\
 & 
	     & & & & 1994 Aug 27 & 49592.6 & 40 \\
 & 
	     & & & & 1994 Oct 13 & 49639.4 & 40 \\
 & 
	     & & & & 1994 Nov 25 & 49682.4 & 30 \\
 & 
	     & & & & 1994 Dec 13 & 49700.3 & 30 \\
 & 
	     & & & & 1995 Jul 22 & 49921.6 & 30 \\
 & 
	     & & & & 1995 Aug 20 & 49950.6 & 30 \\
 & 
	     & & & & 1995 Aug 31 & 49961.5 & 30 \\
 & 
	     & & & & 1996 Oct 1 (E)& 50358.6 & 20x2 \\

I Zwicky 1 & 
	    00 50 57.8 & +12 25 20 & 14.03 & 0.061 & 1991 Oct 16 & 48546.5 & 60 \\
 & 
	     & & & & 1992 Aug 2 & 48837.6 & 30 \\
 & 
	     & & & & 1992 Sep 1 & 48867.0 & 90 \\
 & 
	     & & & & 1993 Aug 12 & 49212.6 & 80 \\
 & 
	     & & & & 1993 Aug 19 & 49219.6 & 60 \\
 & 
	     & & & & 1994 Aug 26 & 49591.6 & 60 \\
 & 
	     & & & & 1994 Dec 13 & 49700.3 & 60 \\
 & 
	     & & & & 1995 Jan 3 & 49721.3 & 30 \\

NAB 0205+024 & 
	    02 05 14.5 & +02 28 42 & 15.40 & 0.155 & 1992 Aug 31 & 48866.6 & 60 \\
 & 
	     & & & & 1992 Sep 1 & 48867.6 & 60 \\
 & 
	     & & & & 1993 Aug 20 & 49220.6 & 90 \\
 & 
	     & & & & 1993 Oct (E)& 49267.2 & 60x2 \\
 & 
	     & & & & 1994 Aug 29 & 49594.6 & 60 \\
 & 
	     & & & & 1994 Sep (E)& 49626.2 & 90x2 \\
 & 
	     & & & & 1994 Oct 13 & 49639.5 & 60 \\
 & 
	     & & & & 1994 Nov 25 & 49682.4 & 60 \\
 & 
	     & & & & 1994 Dec 12 & 49699.4 & 60 \\
 & 
	     & & & & 1995 Jan 2 & 49720.3 & 60 \\

PG 1211+143 &
	    12 11 44.8 & +14 19 53 & 14.63 & 0.085 & 1992 Jan 30 & 48652.6 & 30 \\
 & 
	     & & & & 1993 Feb 11 & 49030.5 & 60 \\
 & 
	     & & & & 1993 Mar 15 & 49062.5 & 40 \\
 & 
	     & & & & 1994 Mar 4 & 49416.5 & 20 \\
 & 
	     & & & & 1994 Dec 13 & 49700.7 & 40 \\
 & 
	     & & & & 1995 Jan 2 & 49720.6 & 60 \\
 & 
	     & & & & 1995 Feb 4 & 49753.6 & 40 \\
 & 
	     & & & & 1995 Mar 18 & 49795.5 & 60 \\
 & 
	     & & & & 1995 Apr 9 & 49817.4 & 40 \\
 & 
	     & & & & 1996 Mar 10 & 50153.5 & 30 \\
 & 
	     & & & & 1996 Mar 11 & 50154.5 & 30 \\

NGC 5548 & 
	    14 15 43.5 & +25 22 01 & 13.73 & 0.017 & 1992 Feb 1 & 48654.6 & 30 \\
 & 
	     & & & & 1992 May 26 & 48769.4 & 30 \\
 & 
	     & & & & 1993 Feb 12 & 49031.5 & 60 \\
 & 
	     & & & & 1993 Mar 14 & 49061.6 & 30 \\
 & 
	     & & & & 1993 Jul 14 & 49183.4 & 30 \\
 & 
	     & & & & 1994 Jan 24 & 49377.7 & 45 \\
 & 
	     & & & & 1994 Mar 4 & 49416.6 & 45 \\
 & 
	     & & & & 1995 Mar 18 & 49795.6 & 30 \\
 & 
	     & & & & 1996 Mar 10 & 50153.6 & 30 \\

Akn 564 & 
	    22 40 18.3 & +29 27 48 & 14.16 & 0.025 & 1992 Aug 1 & 48836.5 & 40 \\
 & 
	     & & & & 1992 Sep 1 & 48867.4 & 30 \\
 & 
	     & & & & 1993 Jul 15 & 49184.6 & 30 \\
 & 
	     & & & & 1993 Aug 10 & 49210.5 & 60 \\
 & 
	     & & & & 1993 Aug 20 & 49220.4 & 60 \\
 & 
	     & & & & 1994 Aug 8 & 49573.5 & 60 \\
 & 
	     & & & & 1994 Aug 26 & 49591.5 & 60 \\
 & 
	     & & & & 1994 Sep (E)& 49625.7 & 30x2+60x1 \\
 & 
	     & & & & 1995 Jul 21 & 49920.6 & 60 \\
 & 
	     & & & & 1995 Aug 29 & 49959.4 & 60 \\
 & 
	     & & & & 1996 Oct 1 (E)& 50358.6 & 20x2 \\

\\
\hline
\hline
\end{tabular}
\end{flushleft}

\end{table*}

The slit width was 2.5 arcsec, matching the typical seeing of the site,
with the exception of the March 1995 observation of NGC 5548, for which
a double width was chosen due to bad seeing conditions, and a few
observations of Akn 564 and Mkn 335, for which the good seeing
ensured that also with a slit 2.0 arcsecs wide all the flux entered the
instrument.

The observations were performed with standard procedures, including the
acquisition of bias and dome flat field frames, for subsequent background
and pixel to pixel sensitivity correction, and of calibration lamp frames for
wavelength calibration;
standard stars were observed to perform flux calibration. 

Together with the data on the NLS1s of our sample, we were
interested in the NGC 5548 observations, to be added to literature data
in the variability comparison as a typical Seyfert 1 galaxy.

The data reduction was performed with
standard IRAF tasks. 
As is usual in variability studies, to
obtain line fluxes which are accurate within a few percent we also 
intercalibrated our spectra to correct for
slit losses, by using the fact that the strong forbidden
lines emitted by the Narrow Line Region remain constant on very long
time-scales, and can therefore be used as standard candles. 
Since the [\ion{S}{ii}] $\lambda$6717.0, $\lambda$6731.3 \AA\ lines
are quite weak and not well resolved in all spectra, we used
the strong [\ion{O}{iii}] $\lambda$4958.9,
$\lambda$5006.8 \AA\ transitions for the calibration in the whole
observed spectral region: as a result, the correction is much more accurate
in the \hb\ portion of the spectrum than in that of \ha, and in
a way not easy to quantify.
The correction parameters (wavelength shift and scale
factor) to be applied to all the
spectra of each object, with reference to one chosen spectrum,
are automatically calculated by a code
(van Groningen \& Wanders \cite{vangron})
which makes use of a $\chi^2$ minimum research procedure on the
difference spectrum in the wavelength range of the 
forbidden lines. We found that after the correction the integrated
fluxes of the two [\ion{O}{iii}] lines for each object differ by less
than 10\% from their average value. 

A possible problem of this method, however, is the 
fact that different seeing conditions 
may cause different portions of the NLR and the BLR to enter the slit 
in observations at different epochs, therefore leading to errors in
the calibration. In Paper I we performed several tests of the data
against this source
of error, concluding that, within the uncertainties intrinsic to the method,
this kind of problem does not appreciably affect the 
reliability of the internal calibration.

To the spectra thus obtained from the Loiano programme,
we added (after having performed on them the usual internal calibration) 
two spectra of Akn 564 and Mkn 335
obtained at the ESO 1.52 m telescope in October 1996
as part of another programme: the covered spectral range is in
this case $\sim$ 2400 \AA\ with a resolution of 4.6 \AA/pixel,
and the observations have been performed with a 2 arcsec slit.
We also included
some of the data from the ESO observations used in Paper I. Table~\ref{log} 
also includes information on these observations, thus giving a
complete list of the data taken by the authors and used throughout
this paper.
 
\section{Variability results and comparison with Seyfert 1's}

The fluxes of the strongest optical
lines (\ha\ and \hb) were measured from all our spectra,
for each object and each observing date, 
by fitting a straight-line continuum under the lines and integrating the
overlying fluxes. 

The uncertainties on these fluxes, and consequently on the
calculated percentage variations, were evaluated as described in
Paper I, the major contribution being the uncertainty on the
internal calibration mentioned above; this contribution was then 
estimated by
slightly perturbing the scale factors of the spectra and measuring the
ranges within which no appreciable changes could be seen.
The obtained percentage uncertainties on the line fluxes fall in
the range 3-5\%, and the errors on the flux variations we used
for the following analysis are therefore approximately twice this
amount.

\begin{figure*}

\resizebox{\hsize}{!}{\includegraphics{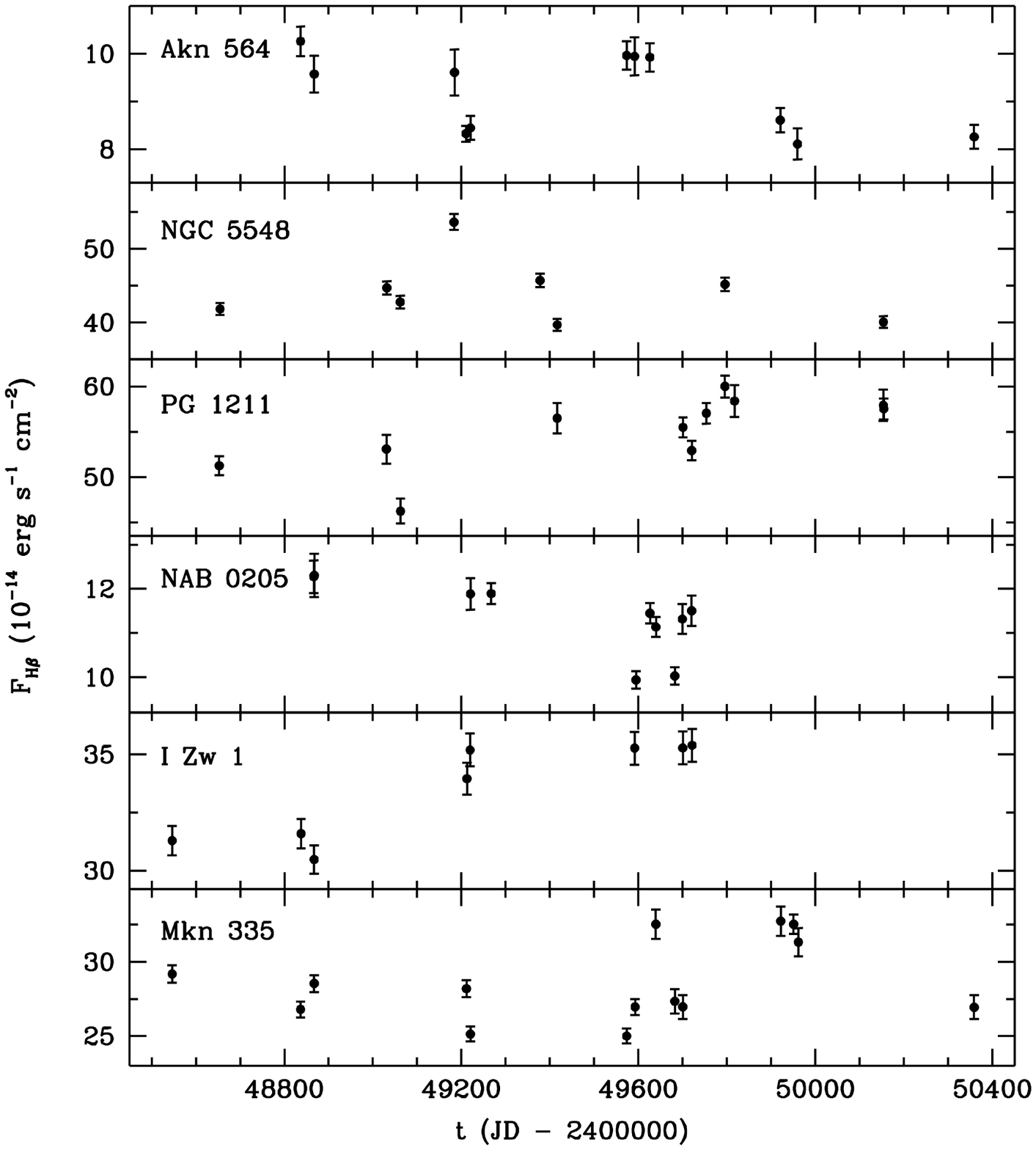}}
\caption{Total \hb\ light curves of the observed NLS1s and of
the Loiano observations of NGC 5548.}
\label{curve}

    \end{figure*}

Figure~\ref{curve} shows the total \hb\ light curves of the observed NLS1s: 
similarly to what we found for the `ESO sample', most objects appreciably
varied their line fluxes at least in some of the considered time 
intervals. The \hb\ light curve of NGC 5548, as given by our Loiano 
observations only, is also plotted in the figure.

In Table~\ref{flussi} the measured \ha\ and \hb\
fluxes for the Loiano and ESO spectra are reported. Not having
enough literature data on the Seyfert 1 
short-term variability of other emission
lines (such as \ion{Fe}{ii} or \ion{He}{ii}), we did not measure their fluxes
and relative variations. We did not evaluate the
continuum flux either, because it is expected to be contaminated by the
host galaxy starlight in different amounts for different observations,
due to the different seeing conditions, slit apertures and source distances;
therefore it is not possible to determine the global continuum
variability properties in a reliable way. 

The variability behaviour of individual Narrow Line Seyfert 1's
in the Loiano (+ ESO) sample is quantified in Table~\ref{var}, 
which reports, for each object, the rms ($\sigma$) of the \hb\ flux 
throughout the whole campaign, and the fractional variation $F_{var}$, 
defined (Clavel et al. \cite{clavel}) as the ratio of the rms fluctuation 
to the mean line flux, corrected for the mean measuring error.
Note that this last quantity is very similar
($\sim$ 5-9\%) for all the NLS1s, while it is much higher for NGC 5548, the only
Seyfert 1 observed at the same telescope and approximately on the
same time baseline, and the data of which are therefore 
also reported in the table for comparison. This is a first indication
of a lesser degree of variability amplitude displayed by Narrow Line
objects.

\begin{table*}
\caption[]{Measured Balmer line fluxes (some points considered less
reliable are indicated in parenthesis).}
\label{flussi}
\begin{flushleft}
\begin{tabular}{lccc|lccc}
\hline
\hline
\\
Galaxy & JD - 2400000 & \ha\footnotemark[1] & \hb\footnotemark[1] &  
Galaxy & JD - 2400000 & \ha\footnotemark[1] & \hb\footnotemark[1] \\
\\
\hline
\\
		
Mkn 335 & 
	     48546.4 & 773 & 292 & 
PG 1211+143 &
	     48652.6 & 1662 & 512 \\
 & 
	     48836.5 & 707 & 268 & 
 & 
	     49030.5 & -- & 531 \\
 & 
	     48867.4 & 817 & 285 & 
 & 
	     49062.5 & 1571 & 462 \\
 & 
	     49211.5 & 884 & 282 & 
 & 
	     49416.5 & 1813 & 565 \\
 & 
	     49220.5 & (471) & 251 & 
 &
	     49700.7 & 1901 & 555 \\
 & 
	     49573.5 & (1355) & 250 & 
 & 
	     49720.6 & 1393 & 529 \\ 
 & 
	     49592.6 & 789 & 270 & 
 & 
	     49753.6 & 1464 & 570 \\
 & 
	     49639.4 & 904 & 325 &  
 & 
	     49795.5 & 1608 & 600 \\
 & 
	     49682.4 & 935 & 273 & 
 & 
	     49817.4 & 1587 & 584 \\
 & 
	     49700.3 & 864 & 270 &  
 & 
	     50153.5 & 1492 & 579 \\
 & 
	     49921.6 & 876 & 327 &  
 & 
	     50154.5 & (2046) & 575 \\
 & 
	     49950.6 & 1090 & 325 &  
NGC 5548 & 
	     48654.6 & 1785 & 418 \\	     
 & 
	     49961.5 & 822 & 313 &  
 & 
             48769.4 & 1098 & 163 \\
 & 
	     50358.6 & 971 & 269 &  
 & 
	     49031.5 & 1998 & 447 \\
 
I Zwicky 1 & 
	     48546.5 & 1096 & 313 & 
 & 
	     49061.6 & 1859 & 428 \\
 &  
	     48837.6 & 1290 & 316 & 
 & 
	     49183.4 & 2234 & 536 \\
 & 
	     48867.0 & 1383 & 305 &  
 & 
	     49377.7 & 1528 & 457 \\
 & 
	     49212.6 & 1295 & 339 &  
 & 
	     49416.6 & 1848 & 397 \\
 & 
	     49219.6 & 1444 & 352 & 
 & 
	     49795.6 & 1862 & 452 \\ 
 & 
	     49591.6 & 1257 & 353 &  
 & 
	     50153.6 & 1858 & 401 \\
 & 
	     49700.3 & 1506 & 353 &  
Akn 564 & 
	     48836.5 & 403 & 103 \\	     
 & 
	     49721.3 & 1448 & 354 &  
 & 
             48867.4 & 415 & 96 \\

NAB 0205+024 & 
	     48866.6 & -- & 123 & 
 & 
	     49184.6 & 368 & 96 \\
 & 
	     48867.6 & -- & 123 &  
 & 
	     49210.5 & 361 & 83 \\
 & 
	     49220.6 & -- & 119 &  
 & 
	     49220.4 & 347 & 84 \\
 & 
	     49267.2 & -- & 119 &  
 & 
	     49573.5 & 389 & 100 \\
 & 
	     49594.6 & -- & 99 &  
 & 
	     49591.5 & 385 & 99 \\
 & 
	     49626.2 & -- & 114 & 
 & 
	     49625.7 & 452 & 99 \\
 & 
	     49639.5 & -- & 111 &  
 & 
	     49920.6 & 341 & 86 \\ 
 & 
	     49682.4 & -- & 100 &  
 & 
	     49959.4 & 410 & 81 \\
 & 
	     49699.4 & -- & 113 &  
 & 
	     50358.6 & 445 & 83 \\
 & 
	     49720.3 & -- & 115 &  
 & 
	     & & \\

\\
\hline
\hline
\end{tabular}
\end{flushleft}
\begin{list}{}{}
\item[$^1$] Units are 10$^{-15}$ erg s$^{-1}$ cm$^{-2}$. 
\end{list}
\end{table*}

To perform a direct quantitative comparison, we first evaluated, 
for each NLS1 in the `Loiano sample', all the relative flux variations 
\footnote{Here and in the following 
we always refer to relative variations
taken from single measurements -- only the 5 ESO points are the result
of averaged spectra -- and computed with respect
to the flux mean value.}
on time intervals ranging from $\sim$ 10 months to $\sim$
14 months; to these data we added the similar ones obtained from
the previous ESO campaign (provided that there are no overlapping
intervals, to avoid to include twice the same relative variation) 
and from the 1996 ESO observations mentioned above. 
We thus obtained a data-set of 28 points (i.e. relative flux
variations) for \hb\ and 27 points for \ha.

\begin{table*}

\caption[]{Global variability properties of the \hb\ flux of each object within the Loiano + ESO monitoring campaigns (see text for details).}
\label{var}

\begin{flushleft}
\begin{tabular}{lccccccccc}
\hline
\hline
\\
Galaxy & & Classification & $\sigma$\footnotemark[1] (F$_{H\beta}$) 
& F$_{var}$ (F$_{H\beta}$) \\
\\
\hline
\\
Mkn 335 & & NLS1 & 27 & 8.9\% \\

I Zwicky 1 & & NLS1 & 21 & 5.4\% \\

NAB 0205+024 & & NLS1 & 8 & 6.6\% \\

PG 1211+143 & & NLS1 & 40 & 6.5\% \\

NGC 5548 & & Sy 1 & 102 & 24.6\% \\

Akn 564 & & NLS1 & 8 & 8.3\% \\

\\
\hline
\hline
\end{tabular}
\end{flushleft}
\begin{list}{}{}
\item[$^1$] Units are 10$^{-15}$ erg s$^{-1}$ cm$^{-2}$.
\end{list}
\end{table*}

As for the Seyfert 1's, also in this case we enriched the 
previous data-set: first we added to the de Ruiter \& Lub Seyfert 1 sample 
(de Ruiter \& Lub \cite{deruiter}, Lub \& de Ruiter \cite{lub};
see also Paper I for details on the sample -- line fluxes have been
obtained in the same way as for ESO and Loiano spectra) the
data of the typical Seyfert 1 NGC 5548 taken from Loiano observations 
for the period since 1992. We then took some points from the
\ha\ and \hb\ NGC 5548 literature data that we collected to perform the
short time-scale comparison (Wamsteker et al. \cite{wamsteker}, Korista
et al. \cite{korista} and references therein).
The data-set includes in this case 135 points for \hb\ and 88 for \ha.

To compare the short-term variability properties of the two classes of
AGNs, we chose a short average time interval, such that the relative
variations measured on this scale would be both representative of the
characteristics of the objects and numerous enough to allow as accurate 
a statistical comparison as possible. Since the Loiano light curves are
not evenly sampled, and the time gaps are sometimes of the order of
several months, we took all the relative variations of the optical lines
on time intervals ranging from $\sim$ 10 days to $\sim$ 45 days, the
average interval being around 27 days. In this way we used all
the data, maximizing the available information and obtaining finally
23 points for \hb\ variations and 17 points for \ha\ ones, having added
to the \hb\ set also 2 variations calculated with ESO Paper I observations.

A similar kind of information can be obtained, in the case of \hb, 
for the typical Seyfert 1 
galaxy NGC 5548, which we take here as representative of the Seyfert 1 class as
regards variability characteristics (although, as mentioned in Section 1, the
generality of its properties is to be determined): 
since in fact this object has been
optically monitored for several years with high temporal resolution, it
is possible to check its monthly variations
by properly sampling its \hb\ light curve for the period $\sim$ 1989-1992
(Korista et al. \cite{korista} and references therein). 
The sampling has been performed by simply taking the observation closest
in time to the epoch which is separated from the previous point by a time
interval equal to the average chosen one (in this way we also `mimick'
the uneven sampling of NLS1 data).
To these observations, sampled so that the average time
interval is of the order of 27 days, we also added the Loiano data
of the same object which were not temporally overlapped to the literature
light curve. We collected in this way a total of 56 points for \hb\ variations.
In the case of \ha, we tried to collect literature
data on several LAG (Robinson \cite{r94}) and AGN Watch monitored objects, 
again sampling the literature
light curves to obtain line flux relative variations on $\sim$ 1 month
time-scale. 
However, this last data-set resulted to be very inhomogeneus; on the
other hand, the [\ion{O}{iii}] correction of the Loiano data does
not give for the \ha\ region a calibration accurate enough to allow
a reliable comparison, and therefore we did not use the \ha\
1-month variations in the following analysis.

\begin{figure*}

\resizebox{\hsize}{!}{\includegraphics{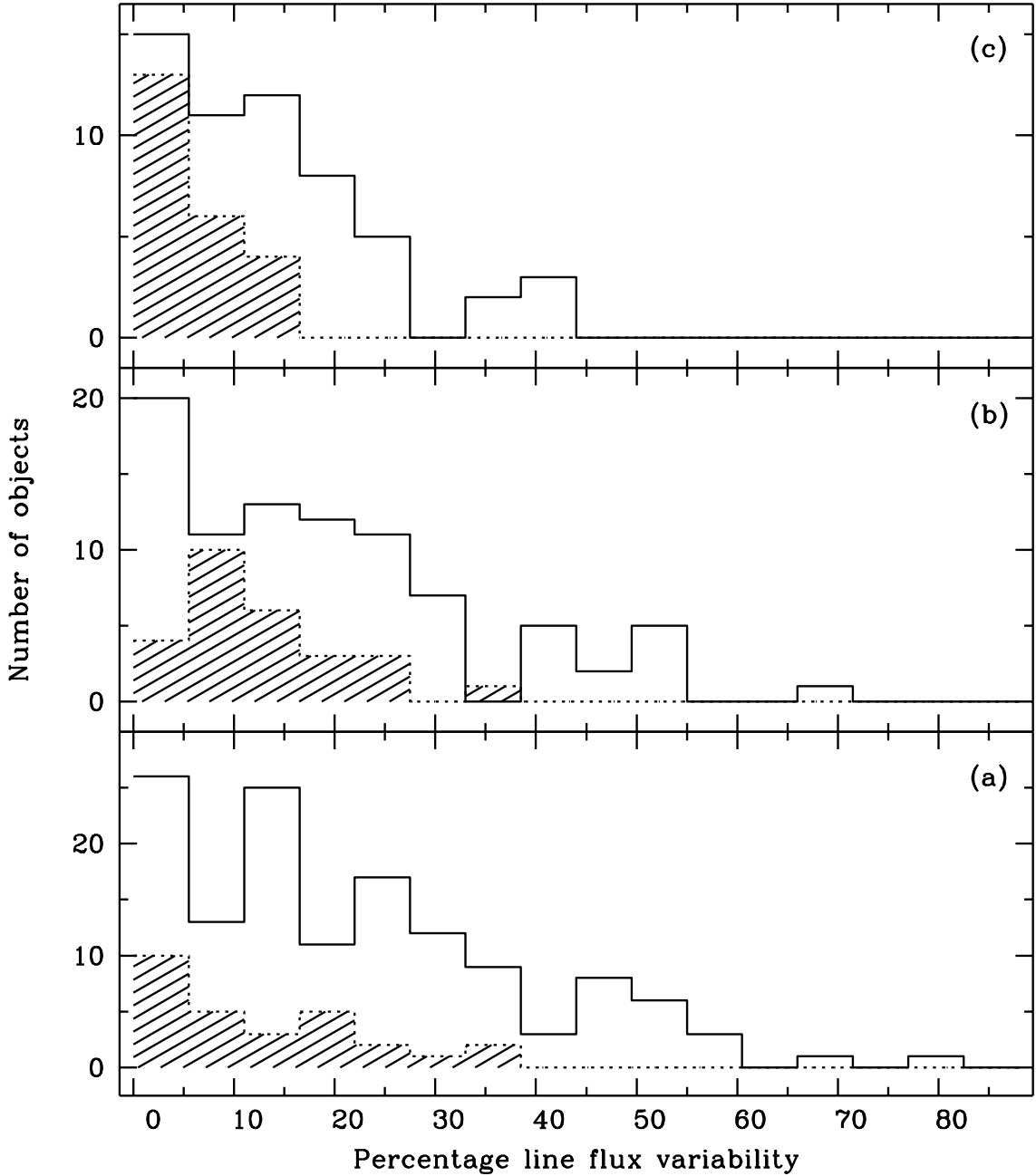}}
\caption{Statistical comparison between the variability properties of 
Seyfert 1 galaxies (open histograms) and NLS1s (shaded
histograms): (a) \hb\ flux variations over a time interval of $\sim$ 1yr;
(b) same as (a) but for \ha\ variations; (c) \hb\ variability on a
monthly time-scale.}
\label{hist}

    \end{figure*}

Similarly to what we did in Paper I, a first statistical 
comparison between Seyfert 1's and NLS1s can be performed with
these data by 
constructing, for each class of object, time-scale of variations, and
emission line, a histogram 
including all the measured relative variations (in
absolute value). We then compared
the resulting histograms relative to Seyfert 1's and NLS1s for each
group of parameters, as plotted in Figure~\ref{hist}a-c: the 1-year data
obtained for \ha\ and \hb\ variations seem now to show a trend of
Narrow Line Seyfert 1's towards weaker variability, and a clearer trend
in this sense is suggested by the 1-month \hb\ results, where the 
NLS1 histograms display a steeper decline and go to zero at around 15-20\%
percentage variability. 

A more quantitative way to evaluate the statistical behaviour of the
two classes of objects is the application of the Kolmogorov-Smirnov
(K-S) test, which gives the probability ($P$) that
two continuous data samples are drawn from the same parent population,
together with the maximum distance ($D$) between the 2 cumulative
probabilities; since $P$ depends on the number of data points in each sample,
larger samples have higher
statistical significance, while more conservative results are obtained 
for smaller samples. This is the case, for example, of the \hb\ line
flux data reported
in Paper I, for which the K-S test gives $D \simeq 0.2$ and 
$P \simeq 58\%$: our conclusion that there was no evidence for a different
behaviour was therefore justified at that time. 

The same test on the
enlarged 1-year samples, however, gives now different results, i.e. 
$D \simeq 0.3$ and $P \simeq 5.4\%$ for \hb\ and
$D \simeq 0.3$ and $P \simeq 3.9\%$ for \ha; so it seems improbable 
that the 2 sets of data are taken from objects with similar
variability properties. A lower probability that their parent
population is the same is obtained by the K-S test applied to \hb\
1-month data: here, in fact, we got
$D \simeq 0.4$ and $P \simeq 0.8\%$. 
Note that neither the histogram appearance nor the K-S test results for
the short-term variations result substantially altered if we sample the
NGC 5548 \hb\ light curve in a slightly different way (e.g. starting
from a different point), the probability $P$ always ranging between
0.5 and 0.9\% and the distance $D$ between 0.41 and 0.43.

\section{Discussion and conclusions}

As already mentioned in Section 1, we discussed the 
various interpretative models for Narrow
Line Seyfert 1 spectra in Paper I, concluding that the possibilities 
of a smaller
black hole mass or a BLR size which is larger on average are the most
probable ones. The other interpretations are in fact in some way
inconsistent with observational results: for example, a projection effect
explanation would imply too narrow a cone of visibility for
NLS1s to justify their being more than 10\% of the known Seyfert 1 population; 
a depletion of the innermost part of the BLR would
produce very low line equivalent widths, while the obscuration hypothesis
is quite unlikely, since the results of X-ray observations all
suggest that we have a direct view of the nucleus.

The knowledge of the variability behaviour of these objects, compared
to other Seyferts, can therefore be crucial in the understanding of their
nature, allowing us to estimate the BLR size. However, the analysis
of variability data can be potentially influenced by some effects:
first of all, if the relative contribution of the (constant) narrow 
component to the total line flux is more important in NLS1s than
in normal Seyfert 1's, then a global weaker variability is observed
in the former objects, even if their broad component fluctuations are
on average of the same amplitude as those of Seyfert 1's. 
However, the evaluation of the ratio between narrow and broad
component of \ha\ and \hb\ in the Stirpe (\cite{stirpe}) Seyfert sample
shows that the relative contributions to the total line fluxes are not
dramatically different in the two classes of objects, the narrow
component being on average $\sim$ 4\% of the broad one in normal Seyfert
1's and $\sim$ 6\% in NLS1s; furthermore, it results to be 6\% in
NGC 5548, which assures that, at least in the short-term comparison
with this object, our measured relative variations are smoothed
by the same factor.
Another possible bias in the interpretation of the results is the unknown
continuum variability which drives the line fluctuations: again, an
intrinsically lower continuum variability amplitude in NLS1s would
result in a weaker line variability, even if the BLR has a `standard'
size in these objects. 
Given the lack of other information (e.g. correlations between optical/UV
variability amplitude and broad line width), we assume that
the ionizing continuum has the same variability properties in these
objects as in other Seyferts.
In both cases, since a more detailed analysis of the data 
(involving the subtraction
of narrow line components and host galaxy continuum from the spectra,
to evaluate the continuum fluctuations)
is beyond the scope of this paper, we will assume that the potential
`smoothing' effects are small compared to the one of which we found evidence.

Based on the ensemble of data that we collected in this work, the
weaker variability of NLS1s, implied by the histogram comparisons and
the K-S test both for long- and short-term observations, seems to exclude
that the broad line emitting gas is located as close to the centre as in
normal Seyfert 1's, as assumed in the `small black hole mass' model.
It seems more probable, instead, that in these objects the gas extends
further out: its distance should be not so high as to completely smear out
the ionizing continuum variations in the line reprocessing, but large
enough to `smooth' their amplitude more than normal BLRs do, producing
in this way the statistically weaker variability that we observe.
This is of course assuming that the physical conditions (like the electron
density or the column density) in the ionized gas emitting the broad
lines are similar in NLS1s and in normal Seyfert 1's.

The fact that we actually observed short-term variations in
these objects, though with small amplitude, implies that,
in the `extended BLR' picture, the
emissivity radius of their BLR gas is
probably closer to our estimate of $r_{min}(NLS1) \sim 3
r_{min}(Sey1)$, made in the case of this model in Paper I, than to the
upper limits that can be set on the basis of the average time-scale
of the observations: the 1-month histograms we constructed include in fact
percentage variations measured on several time intervals, the mean of
which is of the order of 27 days, but which also include $\Delta t$
$\sim$ 10-20 days; the sample is not large enough to allow a more
detailed evaluation of the variability at the level of a few days,
without losing in statistical significance.
Only through an intense monitoring campaign on one or more NLS1s (one
observation every few days) it would be possible
to estimate the BLR size by measuring 
the lag between line and continuum light curves. 

\begin{acknowledgements}

We are grateful to the staff of the Loiano Observatory for the assistance
and to the Loiano Telescope Committee for the generous allocation of observing
time for the monitoring project. We thank Hans de Ruiter and Jan Lub for
allowing us to use their data in advance of publication.

\end{acknowledgements}


\end{document}